\documentstyle[11pt]{article}
\begin{document}
\bibliographystyle{unsrt}
\begin{flushright}
{\normalsize
July 1995}
\end{flushright}
\vskip 0.2in
\begin{center}
{\LARGE {Flavour Equilibration in Quark-Gluon Plasma}}
\vskip 0.2in
{\normalsize Jan-e Alam$^a$, Pradip Roy$^a$, Sourav Sarkar$^a$,
Sibaji Raha$^b$ and Bikash Sinha$^{a,c}$}
\end{center}
\vskip 0.2in
\noindent
{\small {\it a) Variable Energy Cyclotron Centre,
     1/AF Bidhan Nagar, Calcutta 700 064, India}}
\vskip 0.15in
\noindent
{\small {\it b) Bose Institute,
           93/1, A. P. C. Road, Calcutta 700 009,
           India}}
\vskip 0.15in
\noindent
{\small {\it c) Saha Institute of Nuclear Physics,
           1/AF Bidhan Nagar, Calcutta 700 064,
           India}}
\vskip 0.2in
\centerline {\bf {\normalsize Abstract}}
\vskip 0.15in

Within the framework of a dynamical and physically transparent model
developed earlier, we study the time evolution of various quark flavours
in the baryon-free region in ultrarelativistic heavy ion collisions. We
show that even under optimistic conditions, the quark-gluon system fails
to achieve chemical equilibrium.
\vskip 0.2in
\noindent
PACS number(s) : 25.75.+r, 12.38.Mh, 24.60.Ky, 24.85.+p
\vskip 0.25in
\renewcommand{\baselinestretch}{1.5} 
\parindent=20pt

Quantum  Chromodynamics  (QCD), the standard theory of
strong interactions, predicts that at very high temperature
and/or density, the colourless hadronic matter dissolves into
its coloured constituents, the quarks and gluons, such that
the bulk properties of strongly interacting matter are
governed by these degrees  of  freedom.  Such a locally
colour-deconfined phase  of  matter  is  known  as  Quark Gluon
Plasma (QGP). It is expected that the temperature and density
achievable in ultrarelativistic collisions of heavy ions is
favourable for QGP formation, although transiently.

Many aspects of this transition,{\it e.g.}, the order of the
phase transition, the value of the critical temperature,
signals of the transition, thermodynamic equilibration
, are still uncertain. So far,
most of the works assume thermodynamic equilibrium throughout the
evolution history, after an initial proper time  $\tau_i$  ($\sim
1$ fm/c). Thus the thermodynamic quantities like pressure, entropy
and temperature have direct meaning with respect to an equilibrated
system. It has however been recently shown that the approach to
equilibrium in ultrarelativistic heavy ion collisions proceeds
through a succession of many time scales\cite{Jane,Shur}. In
particular, the dominance of the $gg$ cross section over the $qg$
or $qq$ cross sections \cite{Cutler} was argued to imply that the
gluons equilibrate among themsleves appreciably earlier than the
whole system of quarks and gluons. In a recent work \cite{Jane},
referred to as {\bf 1}, we showed that the time scale of {\it
kinetic} equilibration for the light quarks ($u,d$) is $\sim 1$
fm/c; for massive quarks, it increases with the mass to rather
large values. The significance of these considerations for QGP
diagnostics has been discussed in \cite{Raha,Shur2}; the emission
of particles from the pre-equilibrium era ($\tau_g \le \tau  \le
\tau_{th}$), $\tau_{th}$ denoting the time for full thermodynamic
equilibrium, may indeed populate the invariant mass or $p_T$ windows
thought to be relevant for signals from the thermalised QGP.

In this letter, we study the {\it chemical} evolution of the quarks
in the system from the epoch $\tau_g$ ( the proper time when gluons
thermalise) onward. Chemical equilibration has very important
implications for the signal of QGP formation \cite{Raha,Shur2,GK}. It
would be ideal to study this problem in the QCD based kinetic theory and
attempts along these lines have been made recently  \cite{Geig}. Such
calculations, appealing as they are, rely heavily on rather involved
numerical simulations and also suffer from dependence on parameters
needed to mimic the non-perturbative effects. We propose, as in {\bf
1}, a physically transparent scenario which retains the essence of the
kinetic theory to the following extent : the gluons thermalise {\it
completely} at a proper time $\tau_g$ earlier than the quarks. The
gluons carry about half of the total momentum and the quarks carry a
tiny fraction. We restrict ourselves to the  central  rapidity  region
where the number of valence quarks is assumed to be negligibly small.

In  order  to  treat  the  equilibration  of the species one must
follow the microscopic evolution of the phase space  distribution
function  $f(x^\mu,p^\mu)$, governed  by the Boltzmann
transport equation :
\begin{equation}
\hat  L\lbrace
f_q\rbrace  =  \hat C \lbrace f_q\rbrace
\end{equation}
where $\hat L$
is the Liouville operator and $\hat C$ the collision operator,
the subscript $q$ denoting a quark flavour.

The Boltzmann   equation   is   a  non-linear,
integro-differential equation for the  phase  space  distribution
function  of  the  particles. For
our present purpose, however, the problem can be addressed in a
rather simple manner without losing much information on the time
scales for chemical equilibrium.

In  ref. \cite{Early,Early2}  the  decoupling of the relic particles in
the early universe was studied by integrating the Boltzmann
equation over the momentum of the particle to obtain an equation
for the evolution of number density. We use a similar approach to
look at the number density evolution of the partons in
ultrarelativistic collisions of heavy ions; the major difference
between the two cases is that the expansion dynamics in the early
universe ({\it big bang\/}) is governed by the Hubble expansion
\cite{Early2} whereas in the case of heavy ion collision ({\it mini
bang\/}), the expansion dynamics is assumed to be governed by the
boost invariant (Bjorken) solution of realtivistic hydrodynamics
\cite{Bj}. The equation for the number density evolution then reads
\begin{equation}
\frac{dn_q}{d\tau} + \frac{n_q}{\tau}=
\frac{g_q}{(2\pi)^3}\,\int\lbrace\hat C\rbrace
\frac{d^3p}{E}
\end{equation}
where  $n_q$  is  non-equilibrium quark
density, $\tau$ the proper time and $g_q$ the statistical
degeneracy of the quarks of flavour $q$.

The most important contributions to  the  collision term in eq. (2)
are from gluon fusion ($gg\leftrightarrow q\bar q$) and gluon decay
($g\rightarrow q\bar q$) \cite{Alth1,Alth2}. Since we have assumed
complete equilibration for the gluons, reactions like $gg\rightarrow
ggg...$ are not included in our work, the gluon density being
determined by the temperature of the system at each instant. We
ignore the quantum effects but take into account relativistic
effects because of the very high temperature. Thus gluons are
described by relativistic  Maxwell  -  Boltzmann statistics,
$f_g(E)=g/{(2\pi)^3}e^{-E/T}$. Then eq.(2) can be written as,

\begin{equation}
\frac{dn_q}{d\tau}
=-\frac{n_q}{\tau}-R_{gg\rightarrow q\bar q}(T)
\frac{(n_q^2-n_{eq}^2)}{n_{eq}^2}+R_{g\rightarrow q\bar q}(T)
\end{equation}
where  $n_{eq}(T)=g_q/(2\pi)^3\int{d^3p/(e^{E/T}+1)}$, is
the equilibrium density. Eq.(3) is a particular form of the Riccati
equation; there is no closed form solution of this equation.

In deriving eq.(3) we have assumed T (or CP) invariance, {\it i.e.},
$|M|_{gg\rightarrow q\bar q}=|M|_{q\bar q\rightarrow gg}$.
$R_{gg\rightarrow q\bar q}(T)$ is the quark  production  rate  per
unit four volume by the reaction $gg\rightarrow  q\bar q$ and
$R_{g\rightarrow q\bar q}$ gives the same for the gluon decay,
$g\rightarrow q\bar q$ \cite{Alth1,Alth2}. The thermal quark masses
have been taken into account \cite{Jane} using $m_q^2(T)=m_q^2+g^2T^2/4.5$
in all the rates which was neglected in the work of Bir\'o {\it et al}
\cite{Biro}. The full derivation of the above equation has been omitted
here for the sake of brevity; it will be reported elsewhere \cite{tobe}.
Eq.(3)  is a very convenient form of the Boltzmann transport
equation with all the relevant features for our
present purpose. The first term on the right hand side of the equation
represents the "dilution" of the density due to one dimensional
expansion, the  destruction of the quarks is propotional to the
annihilation rates of $q\bar q$ and destruction is balanced by creation
through the inverse reaction when $n_q=n_{eq}$. The last term stands for
the production of quarks due to gluon decay for which there is clearly no
inverse process.

To solve eq.(3) for $n_q(\tau)$ we need to know $T(\tau)$, which
is obtained by solving the energy conservation equation
\begin{equation}
\frac{d\epsilon}{d\tau}+\frac{\epsilon+P}{\tau}=0
\end{equation}
where $P$ is the pressure and $\epsilon$ the energy density. An
equation of state relating $P$ and $\epsilon$ can be parametrized in
the form
\begin{equation}
P=c_{s}^{2}\epsilon
\end{equation}
Then, for the non-equilibrium configuration \cite{Note}, we can write
\begin{equation}
\epsilon=\left[ar_g+br_q(\tau)\right]T^4
\end{equation}
where $a=(8\pi^2)/45\,c_{s}^{-2}$ and
$b=(7\pi^2)/60\,N_f\,c_{s}^{-2}$. In chemical equilibrium,
$r_q (\equiv n_q/n_{eq})$ for all species should be unity
and in correspondence with our premise, we take $r_g$ =1
all through. ${N_f}$ is the {\it effective} number of
{\it massless} flavours; appearance of quark masses  amounts to
reducing the actual number of flavours \cite{Hwa,Somenath}. For the
present, we confine our attention only to the three lightest flavours,
$u,d$ and $s$ \cite{Note2}.

We then have,
\begin{equation}
\frac{dT}{d\tau}=\frac{bT}{4(an_{eq}+bn_q)}\,\left[\frac{n_q}{\tau}+
R_{gg\leftrightarrow q\bar q}(T)
\frac{(n_q^{2}-n_{eq}^{2})}{n_q^{2}}\,-\,R_{g\rightarrow q\bar q}(T)\right]
-\frac{T}{3\tau}
\end{equation}
To  solve  eq.(3), we need to specify the initial conditions. The
initial  values  of  $n_q(\tau_g)$'s  are   obtained   from   the
integration    of    the    structure   function \cite{Gluck},
\begin{equation}
n_q=2A\times\frac{\int_{x_{min}}^1u_q(x)dx}{\pi R_A^2\tau_g}
\end{equation}
where  $A$  is the  mass  number  of  the  colliding  nuclei  (208 in
our case for Pb), $x_{min}$ is the minimum value of  $x$, taken to be
0.02 \cite{Jane,Shur}, dictated by the applicability of perturbative QCD.
The values of $n_q(\tau_g)=1.42, 2.54$ and $3.05$ for SPS, RHIC and LHC
respectively. The corresponding initial values  of  the  temperature
($T_g$) are 330, 500 and 660 MeV at proper times $\tau_g$=$0.54$,
$0.3$ and $0.25$ fm/c, respectively \cite{Shur}.

The solutions of eqs.(3) and (7), obtained by the Runge-Kutta method,
provide us with the non-equilibrium density ${n_q(\tau)}$ and the cooling
law,
\begin{equation}
T^3\tau=\left[\frac{a+br_q(\tau_g)}{a+br_q(\tau)}
\right]^{3/4} T_g^3\tau_g
\end{equation}
As is evident from the above discussions, the non-equilibrium density
$n_q$ has an explicit dependence on $\tau$ and an implicit dependence
on $\tau$ through $T(\tau)$. But the equilibrium density $n_{eq}(T)$
has only an implicit dependence on $\tau$ through $T(\tau)$. The
ratio $r_q$ thus assumes an {\it universal} feature, since the implicit
time dependence gets eliminated. The time dependence of the ratio $r_q$ can
then be used as a ready marker for chemical equilibrium; the time at which
the explicit time dependence of ${r_q}$ vanishes corresponds to the time
for chemical equilibration for the flavour $q$.

In contrast to the earlier work of Bir\'o {\it et al} \cite{Biro}, we have
not included reactions like ${q\bar q\rightarrow Q\bar Q}$, since
the initial system is dominated by gluons\cite{Shur,Geig};
the quark density is very low compared to the gluons. These
reactions are suppressed by a factor of about ${1/16}$
\cite{Shur2} compared to the case when the quarks are in complete equilibrium.
Bir\'o {\it et al} \cite{Biro} also estimated their initial density from the
HIJING model \cite{Wang}.

The ratios $r_q=n_q/n_{eq}$ are plotted in fig.1 for SPS, RHIC and LHC
energies. At RHIC and LHC energies, the ratios for $u(d)$ quarks and $s$
quarks are the same at early times but at late times the $u(d)$ quarks
dominate over the $s$ quarks. The reason is that at early times the
thermal mass $(\propto T)$ dominates over the current quark mass but
at later times the thermal mass becomes small and the main contribution
comes from the current quark mass which is large for $s$ quarks
$(m_s \sim 150 MeV)$ compared to the $u(d)$ quark masses $(m_{u(d)}
\sim 10 MeV)$; there occurs thus some Boltzmann suppression for $s$ quarks
relative to $u(d)$ quarks. At SPS energies, at early times the
ratio for $s$ quarks is large compared to the $u(d)$ quarks, this is due
to the initial normalisation. The non-equilibrium density
$n_{u(d)}\sim n_s$ but the equilibrium density for $s$ quarks
is small compared to $u$ and $d$ quarks due to larger mass.

At small time scales (larger temperature) the slope of the ratio for
LHC energies is opposite to that of SPS and RHIC. The non-equilibrium
density contains the effect of expansion as well as production, where
as the equilibrium density is affected by expansion only. In the ratio,
the effect of production dominates over the expansion at LHC i.e. larger
temperature, so the ratio increases. However at RHIC and especially
at SPS the quark production is not so intense to dominate over expansion.

In fig.1, we have also shown the time variation of temperature, given by
the cooling law (eq.(9). The lifetime of the quark-gluon system is
the time when $T$ falls below $T_c$, taken to be 160 MeV as an optimistic
estimate. It is abundantly clear that under either of the two criteria,
$r_q \rightarrow$ 1 and/or flatness of $r_q$ - the quarks remain
out of chemical equilibrium in the baryon-free region at SPS, RHIC and
LHC energies. Detailed account of these effects in QGP diagnostics must
be considered; work along this line is in progress.

\newpage
\section*{Figure Captions}

\noindent Figure 1: Ratio of non-equilibrium density to equilibrium
density and temperature as functions of time at SPS, RHIC and LHC energies.


\begin{thebibliography}{99}
\bibitem{Note} The medium effects enter through the thermal mass of the
quarks; the higher the temperature, the higher the thermal mass. Therefore
the equation of state is parametrized as $P=c_{s}^{2}(T)\epsilon$, where
$c_{s}^{-2}=\left[\frac{T}{g_{eff}(T)}\,\frac{dg_{eff}}{dT} + 3\right]$
with $g_{eff}(T) = (\epsilon+P)/(2\pi^2 T^4)/45$. This would amount to
making $a$ and $b$ in eq. (6) functions of $T$. For the cases considered
here, such effects are negligibly small.
\end{thebibliography}
\end{document}